\documentclass[12pt]{iopart}
\usepackage{graphicx}

\begin{document}

\title{Possible interpretation of the $Z_b$(10610) and $Z_b$(10650) in a chiral quark model}

\author{You-Chang Yang$^{1,2}$, Jialun Ping$^2$ \footnote{corresponding author},
Chengrong Deng$^{3}$ and Hong-Shi Zong$^{4}$}

\address{$^1$Department of Physics, Zunyi Normal College, Zunyi
563002, People's Republic of China \\
$^2$Department of Physics, Nanjing Normal University, Nanjing 210097,
People's Republic of China \\
$^3$School of Mathematics and Physics, Chongqing Jiaotong
University, Chongqing 400074, People's Republic of China \\
$^4$Department of Physics, Nanjing University, Nanjing 210093,
People's Republic of China}
\ead{jlping@njnu.edu.cn}

\begin{abstract}
Motivated by the two charged bottomonium-like resonances
$Z_b$(10610) and $Z_b$(10650) newly observed by the Belle
collaboration, the possible molecular states composed of a pair of
heavy mesons, $B\bar{B},~B\bar{B}^*,~B^*\bar{B}^*,~B_s\bar{B}$, etc
(in S-wave), are investigated in the framework
of chiral quark models by the Gaussian expansion method. The bound
states $B\bar{B}^*$ and $B^*\bar{B}^*$ with quantum numbers
$I(J^{PC})=1(1^{+-})$, which are good candidates for the
$Z_b(10610)$ and $Z_b(10650)$ respectively, are obtained. Other
three bound states $B\bar{B}^*$ with $I(J^{PC})=0(1^{++})$,~
$B^*\bar{B}^*$ with $I(J^{PC})=1(0^{++}),~0(2^{++})$ are predicted.
These states may be observed in open-bottom or hidden-bottom decay
channel of highly excited $\Upsilon$. When extending directly the
quark model to the hidden color channel of the multi-quark system,
more deeply bound states are found. Future experimental search of
those states will cast doubt on the
validity of applying the chiral constituent quark model to the
hidden color channel directly.

\end{abstract}
\submitto{J. Phys. G}
\maketitle

\section{Introduction}
Very recently, the Belle collaboration observed two narrow peaks,
which are named $Z_b$(10610) and $Z_b$(10650), in the
$\pi^{\pm}\Upsilon(nS)~(n=1,2,3)$ and $\pi^{\pm}h_b(mP)~(m=1,2)$
invariant mass spectra in the hidden-bottom decay channels of
$\Upsilon(5S)$~\cite{Zb(10610)}. The measured masses and widths of
the two structures are,
\begin{eqnarray}
&&M_{Z_b(10610)}=10608.4\pm 2.0~ \mbox{MeV},~\Gamma=15.6 \pm 2.5 ~\mbox{MeV}\nonumber\\
&&M_{Z_b(10650)}=10653.2\pm 1.5~ \mbox{MeV},~\Gamma=14.4 \pm 3.2~
\mbox{MeV}.\nonumber
\end{eqnarray}
Analysis favors quantum numbers of $I^G(J^P)=1^+(1^+)$ for both
states. The $Z_b$(10610) and $Z_b$(10650) are both charged
bottomonium-like resonances and the masses are very close to the
thresholds of the open bottom channels $B^*\bar{B}$(10604.6 MeV) and
$B^*\bar{B}^*$ (10650.2 MeV), so the molecular states of S-wave
$B^*\bar{B}$ and $B^*\bar{B}^*$ assignment are suggested by Belle
collaboration.

In the hadron level, Nils A. T\"{o}rnqvist investigated the
deuteron-like meson-meson bound states by meson exchange
model~\cite{zpc61523}. The study shows that the energy of isoscalars
$B\bar{B}^*$ with $J^{PC}=0^{-+},1^{++}$,~ $B^*\bar{B}^*$ with
$J^{PC}=0^{++},0^{-+},1^{+-},2^{++}$ are about 50 MeV below the
corresponding $B\bar{B}^*$ and $B^*\bar{B}^*$ thresholds. No bound
state, however, appears for isovectors. Recently, by taking the
pseudoscalar, scalar and vector mesons exchange into account in the
framework of the meson exchange model, Liu \textit{et al}. found
that the loosely bound states probably exists in S-wave $B\bar{B}^*$
\cite{epjc5663,epjc61411}. Very recently, Sun \textit{et al}.
believe that the $Z_b$(10610), $Z_b$(10650) are respectively
$B^*\bar{B}$ and $B^*\bar{B}^*$ molecular state after considering
$S$-wave and $D$-wave mixing \cite{Sun:2011uh,arXiv:1111.2921}.

In the quark level, by solving the resonating group method equation,
Liu \textit{et al}.~\cite{PRC80015208} also investigated the system
composed of $[\bar{b}q][b\bar{q}]$, ~$[\bar{b}q]^*[b\bar{q}]$,
$[\bar{b}q]^*[b\bar{q}]^*~(q=u,d,s)$ by two chiral quark models in
which the pseudoscalar, scalar and vector mesons exchange are taken.
The isoscalars $B\bar{B}$, $B\bar{B}^*(C=+)$, $B^*\bar{B}^*(J=2)$
favor molecular states. Bondar \textit{et al}. also discussed the
heavy quark spin structure of the $Z_b$(10610) and $Z_b$(10650)
assuming that these are molecular state $B^*\bar{B}$ and
$B^*\bar{B}^*$ \cite{arXiv:1105.4473}. By considering the
contribution from the intermediate $Z_b$(10610) and $Z_b$(10650)
states to the $\Upsilon(5S)\rightarrow \Upsilon(2S) \pi^+\pi^-$
decay process, the anomalous $\Upsilon(2S)\pi^+\pi^-$ production
near the peak of $\Upsilon(5S)$ at $\sqrt{s}=10.87$
GeV~\cite{prl100112001}, observed by Belle collaboration, can be
explained naturally~\cite{arXiv:1105.5193}. The possibility of
$Z_b(10610)$ and $Z_b(10650)$ being tetraquark states are discussed
by the authors of Ref \cite{PhysRevD.85.054011,arXiv:1106.2284}. The
authors of Ref \cite{PLB704312,PhysRevD.85.074014}. think the
tetraquark and molecular structure both can interpret the
$Z_b(10610)$ and $Z_b(10650)$ in the QCD sum rule calculation.
Further theoretical efforts concern the decay and mass of the
$Z_b(10610)$ and $Z_b(10650)$ states discussed in Refs.
\cite{Cleven:2011gp,arXiv:1203.1894,arXiv:1204.3959}.

Inspired by the new states $Z_b$(10610) and $Z_b$(10650) reported by
Belle collaboration~\cite{Zb(10610)} and the related work, a
systematical study of the
possible S-wave $\mathcal{B}\bar{\mathcal{B}}$,
~$\mathcal{B}\bar{\mathcal{B}}^*$ and
${\mathcal{B}}^*\bar{\mathcal{B}}^*$ states is performed in this
work. Here the $\mathcal{B}$ and $\bar{\mathcal{B}}$ stand for
$(B^{+},B^0,B_s^0)$ and $(B^{-},\bar B^0,\bar B_s^0)$ triplets,
respectively. It is worthwhile to investigate the intrinsic
structure of the $Z_b$(10610), $Z_b$(10650) and other possible
exotic states with $b,\bar{b}$ quarks, especially in view of the
great potential of finding new particles at Belle, BaBar, LHC and
other collaborations.

To study the mass spectrum of above possible exotic states, two types
of chiral quark models (ChQM) \cite{slamanca} are employed in this work. 
The numerical method, which is able to provide almost exact
solutions, is very important in the study of few-body systems.
Here, a high precision numerical method for few body system, which
is different from the methods used in the previous work by
other researchers, the Gaussian Expansion Method (GEM) is used.
The detail of GEM can be found in Refs.~\cite{GEM,80114023}.

The paper is organized as follows. In the next section we introduce
the Hamiltonian of the chiral quark models. Section \ref{WF} is
devoted to discuss the wave function of possible molecular states
$\mathcal{B}\bar{\mathcal{B}}$,~$\mathcal{B}\bar{\mathcal{B}}^*$ and
$\mathcal{B}^*\bar{\mathcal{B}}^*$. In Section \ref{result}, we
present and analyze the results obtained in our calculation.
Finally, the summary of the present work is given in the last
section.

\section{The chiral constituent quark model}
In the ChQM, the Hamiltonian usually includes Goldstone-boson
exchange in addition to color confinement and one-gluon-exchange (OGE).
The chiral partner, $\sigma$-meson, is also usually introduced,
although its existence is still in controversy~\cite{sigma}. The
Hamiltonian of the ChQM used here is given as follows,
\begin{equation}
H=\sum_{i=1}^4 \left( m_i+\frac{\mathbf{p}_i^2}{2m_i}
\right)-T_{CM} +\sum_{j>i=1}^4
(V_{ij}^C+V_{ij}^G+V_{ij}^\chi+V_{ij}^\sigma),
\label{h_cqm}
\end{equation}
where $\chi=\pi,K,\eta$, $T_{CM}$ is the kinetic energy operator of the
center-of-mass motion of whole system.

The linear confining potential, which is suggested by lattice QCD
calculation of $q\bar{q}$ system, can be written as

\begin{equation}
V_{ij}^{C} ={\boldmath{\mbox{$\lambda$}}}^c_{i} \cdot
{\boldmath{\mbox{$\lambda$}}}^c_{j}~(-a_{c}r_{ij}-
\Delta).~\label{confinement}
\end{equation}

For one-gluon-exchange, the potential takes the form
\begin{eqnarray}
V_{ij}^G =\alpha_{s} \frac{{\boldmath{\mbox{$\lambda$}}^{c}_{i}\cdot
{\boldmath{\mbox{$\lambda$}}}^{c}_{j}}}{4} \left[{\frac{1}{r_{ij}}}-
{\frac{2\pi}{3m_im_j}}~({\boldmath{\mbox{$\sigma$}}}_{i}\cdot
{\boldmath{\mbox{$\sigma$}}}_{j})~\delta(\mathbf{r}_{ij})
 \right],  \label{gluon}
\end{eqnarray}
where, $\boldmath{\mbox{$\sigma$}}$, $\boldmath{\mbox{$\lambda$}}$ are the
SU(2) Pauli matrices and the SU(3) Gell-Mann matrices, respectively.
The $\boldmath{\mbox{$\lambda$}}$ should be replaced by
$-\boldmath{\mbox{$\lambda$}}^*$ for the antiquark.
In the non-relativistic quark model, the delta function
$\delta(\mathbf{r}_{ij})$ should be regularized \cite{bhad}, because of
the finite size of the constituent quark. The regulation is flavor
dependent and reads~\cite{slamanca,weinstein}
\begin{equation}
\delta(\mathbf{r}_{ij})=\frac{1}{4\pi r_{ij}~r_0^2(\mu)}~
e^{-r_{ij}/r_0(\mu)},
\end{equation}
where $r_0(\mu)=r_0/\mu$ and $\mu$ is the reduced mass of
quark-quark (or antiquark) system.
The wide energy covered from light to heavy quark requires an
effective scale-dependent strong coupling constant $\alpha_s$ in
Eq. (\ref{gluon}) that cannot be obtained from the usual one-loop
expression of the running coupling constant because it diverges when
$Q\rightarrow\Lambda_{QCD}$. Hence an effective scale-dependent
strong coupling constant \cite{slamanca} is taken as
\begin{equation}
\alpha_s(\mu)=\frac{\alpha_0}{\ln\left[\frac{\mu^{2}+\mu_0^2}{\Lambda_0^2}\right]}~,
\end{equation}
where $\mu_0$ and $\Lambda_0$ are the free parameters.

For the mesons exchange, potential takes the form
\begin{equation}
V_{ij}^{\pi}
=C(g_{ch},\Lambda_{\pi},m_{\pi}){\frac{m_{\pi}^{2}}{{12m_{i}m_{j}}}}
H_1(m_{\pi},\Lambda_{\pi},r_{ij})
({\boldmath{\mbox{$\sigma$}}}_{i}\cdot
{\boldmath{\mbox{$\sigma$}}}_{j})~\sum_{a=1}^3\lambda_{i}^{a}\cdot
\lambda_{j}^{a}, \label{pi}
\end{equation}
\begin{equation}
V_{ij}^{K}=C(g_{ch},\Lambda_{K},m_{K}){\frac{m_K^{2}}{{\
12m_{i}m_{j}}}} H_1(m_K,\Lambda_K,r_{ij})
({\boldmath{\mbox{$\sigma$}}}_{i}\cdot
{\boldmath{\mbox{$\sigma$}}}_{j})~\sum_{a=4}^7\lambda_{i}^{a}\cdot
\lambda_{j}^{a}, \label{k}
\end{equation}
\begin{eqnarray}
V_{ij}^{\eta}
=&&C(g_{ch},\Lambda_{\eta},m_{\eta}){\frac{m_{\eta}^{2}}{{
12m_{i}m_{j}}}} H_1(m_{\eta},\Lambda_{\eta},r_{ij})
({\boldmath{\mbox{$\sigma$}}}_{i}\cdot
{\boldmath{\mbox{$\sigma$}}}_{j})~\nonumber\\&&
\times\left[\cos\theta_P(\lambda_{i}^{8}\cdot
\lambda_{j}^{8})-\sin\theta_P(\lambda_{i}^{0}\cdot \lambda_{j}^{0})
\right],\label{eta}
\end{eqnarray}
\begin{equation}
V_{ij}^{\sigma} = -C(g_{ch},\Lambda_{\sigma},m_{\sigma})~
H_2(m_{\sigma},\Lambda_{\sigma},r_{ij}) \label{sigma}
\end{equation}
\begin{equation}
H_1(m,\Lambda,r)=\left[ Y(mr)-{\frac{\Lambda^{3}}{m^{3}}} Y(\Lambda
r)\right]
\end{equation}
\begin{equation}
H_2(m,\Lambda,r)=\left[ Y(mr)-{\frac{\Lambda}{m}} Y(\Lambda
r)\right]
\end{equation}
\begin{equation}
C(g_{ch},\Lambda,m)={\frac{g_{ch}^{2}}{{4\pi
}}}{\frac{\Lambda^{2}}{{\Lambda^{2}-m^{2}}}} m
\end{equation}
where the $\sigma$ exchange only occurs between the lightest quarks
($u$- or $d$-quark) which is different from Ref. \cite{slamanca} due to
its non-strange nature. The strange scalar meson (with large mass) exchange
is not taken into account in the present work because of its small effect.
We adopt $\lambda^{0}=\sqrt{\frac{2}{3}}I$ due to the normalization of
SU(3) matrix. $Y(x)$ is the standard Yukawa function defined by
$Y(x)=e^{-x}/x$ and the rest symbols have their usual meaning. The
chiral coupling constant $g_{ch}$ is determined from the $\pi NN$
coupling constant through
\begin{equation}
\frac{g_{ch}^{2}}{4\pi }=\left( \frac{3}{5}\right) ^{2}{\frac{g_{\pi NN}^{2}%
}{{4\pi }}}{\frac{m_{u,d}^{2}}{m_{N}^{2}}},
\end{equation}
and flavor $SU(3)$ symmetry is assumed. The tensor term and the
spin-orbital term have been omitted in the potentials since we
consider only S-wave states.

The above model is denoted as ChQM1.
To testing the effect of $\sigma$-exchange between the lightest and
strange quark or strange quark pairs, and long-range color screening
on the binding energy of the molecular states, the Salamanca version of
the chiral quark model \cite{slamanca}, which is referred as ChQM2,
is also employed here. The screened confinement interaction in this model is
\begin{equation}
V_{ij}^{C} ={\boldmath{\mbox{$\lambda$}}}^c_{i} \cdot
{\boldmath{\mbox{$\lambda$}}}^c_{j}~\{-a_{c}(1-e^{-\mu_c r_{ij}})+
\Delta \},~\label{confinement2}
\end{equation}
where $\mu_c$ is a color screening parameter. The other potentials are the same
as the above with the exception that the $\sigma$-meson is exchanged between all
the light quarks, $u,d,s$.

\section{Wave function \label{WF}}
The total wave function of multi-quark system can be written as,
\begin{eqnarray}
\Psi^{I,I_z}_{J,J_z} & = & \left|\xi\right\rangle
\left|\eta\right\rangle^{II_z}\Phi_{JJ_z},\label{twave} \\
\Phi_{JJ_z} & =& \left[\left|\chi\right\rangle_{S}
 \otimes\left|\Phi\right\rangle_{L_T}\right]_{JJ_z} \nonumber
\end{eqnarray}
where $\left|\xi\right\rangle$, $\left|\eta\right\rangle^{I}$,
$\left|\chi\right\rangle_{S}$, $\left|\Phi\right\rangle_{L_{T}}$
represent color singlet, isospin with $I$, spin with $S$ and spacial
wave function with angular momentum $L_T$, respectively.

All possible molecule structures composed of S-wave $\mathcal{B}$
and $\bar{\mathcal{B}}$, which stand for $(B^{+},B^0,B_s^0)$ and
$(B^{-},\bar B^0,\bar B_s^0)$ triplets, respectively, are
investigated in this work. According to the total isospin, the
$P\bar P$ (pseudoscalar meson) and $V \bar V$ (vector meson) flavor
wave functions are listed in Table \ref{PPPV}. Another possible
molecule structure for four-quark system is bottomonium+light meson.
In this case, there is no interaction between colorless bottomonium
and light meson (color dependent interaction is zero between two
colorless cluster if no exchange term exists and there is also no
Goldstone-boson exchange between heavy and light quarks), so no
bound state can be formed in this case. Therefore, we do not take
into account of this case in the present work.

Obviously the components $P \bar V$ and $V \bar P$ do not have
definite $C$ parity, one can get $C$ parity $=\pm$ by $[P \bar V \pm
\hat C\,(P \bar V)]/\sqrt{2}$  and  $[V \bar P \pm \hat C\,(V \bar
P)]/\sqrt{2}$ for the neutral states \cite{zpc61523}  such as
$(\mathbf{I},I_z)=(1,0),~(0,0)$ shown in Table \ref{PPPV} (All the
orbital angular momenta are set to zero because we concentrate on
ground states). Hence, the coefficient $C=\pm1$ represent $C$-even
and -odd parity respectively, which are different from that of
Ref.~\cite{epjc61411,epjc5663,liu-4430}, since we use normal
convention of PDG~\cite{PDG} i.e. $B^0=d\bar{b}$ and
$\bar{B}^0=\bar{d}b$. One can easy find that $G$ parity $=\mp$ are
corresponding to $C$ parity $=\pm$ for these states with
$(\mathbf{I},I_z)=(1,0)$. However, there is no interaction depending
on $C$ and $G$ parity in the model Hamiltonian Eq.(\ref{h_cqm}), so
these two states with $\pm$ $C$ and $G$ parity must be degenerate in
our calculation. The two states separated by comma in each row of
Table \ref{PPPV} will be coupled in the calculation.
\begin{table}[htb]
\caption{The flavor wave functions of the
$\mathcal{B}\bar{\mathcal{B}}$, $\mathcal{B}\bar{\mathcal{B}}^*$ systems.
``$C$" is the charge parity. \label{PPPV}}
\begin{indented}
\lineup
\item[]\begin{tabular}{@{}lcc}
\br
Isospin           & $\mathcal{B}\bar{\mathcal{B}}$ & $\mathcal{B}\bar{\mathcal{B}}^*$ \\ \mr
                  & $B^+\bar{B}_{s}^0$ & $B^{*+}\bar{B}_{s}^0,~B^{+}\bar{B}_{s}^{*0}$ \\
I=$\frac{1}{2}$   & ${B}^0 \bar B_s^0$ & $B^{*0}\bar{B}_{s}^0,~B^{0}\bar{B}_{s}^{*0}$ \\
                  & ${B}_s^0 \bar B^0$ & ${B}_s^{*0} \bar B^0,~{B}_s^{0} \bar B^{*0}$ \\
                  & ${B}_s^0 B^-$ & ${B}_s^{*0} B^-,~{B}_s^0 B^{*-}$ \\ \mr
                  & $B^+\bar B^0$ & $B^{*+}\bar{B}^0,~ B^{+}\bar{B}^{*0}$ \\
I=1               & $\frac{1}{\sqrt{2}}(B^+B^--B^0\bar B^0)$ &
                    $\frac{1}{2}[(B^{*+}B^--B^{*0}\bar B^0)+C\,(B^{*-}B^+ -\bar B^{*0}B^0)]$, \\
                  & & $\frac{1}{2}[(B^+B^{*-}-B^0\bar B^{*0})+C\,(B^-B^{*+}-\bar B^0 B^{*0})]$ \\
                  & ${B}^0 B^-$ & ${B}^{*0} B^-,~ {B}^0 B^{*-}$ \\ \mr
I=0($l$)          & $\frac{1}{\sqrt{2}}(B^+B^-+B^0\bar B^0)$ &
                    $\frac{1}{2}[(B^{*+}B^-+B^{*0}\bar B^0)+C\,(B^{*-}B^+ +\bar B^{*0}B^0)]$,\\
                  & & $\frac{1}{2}[(B^+B^{*-}+B^0\bar B^{*0})+C\,(B^-B^{*+}+\bar B^0 B^{*0})]$ \\
I=0($s$)          & $B_s^0 \bar{B}_s^0$ & $\frac{1}{\sqrt{2}}(B_s^{*0} \bar B_s^0+C\, \bar B_s^{*0}
               B_s^0),~\frac{1}{\sqrt{2}}(B_s^0 \bar B_s^{*0}+\bar B_s^0 B_s^{*0})$ \\ \br
\end{tabular}
\end{indented}
\end{table}

\begin{figure}[htb]
\begin{center}
\includegraphics{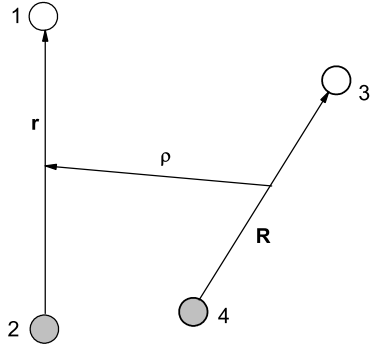}
\end{center}
\vspace*{-0.5cm}
\caption{The relative coordinate for a meson and antimeson system.
solid and hollow circles represent quarks and antiquarks,
respectively.} \label{jacobi}
\end{figure}

The spatial structures of molecular states are pictured in Fig.
\ref{jacobi}. The relative coordinates are defined as following,
\begin{eqnarray}
&&\mathbf{r}=\mathbf{r}_1-\mathbf{r}_2,\ ~~~~ \mathbf{R}=\mathbf{r}_3-\mathbf{r}_4,\\
&&\mathbf{\rho}=\frac{m_1\mathbf{r}_1+m_2\mathbf{r}_2}{m_1+m_2}
-\frac{m_3\mathbf{r}_3+m_4\mathbf{r}_4}{m_3+m_4},
\end{eqnarray}
and the coordinate of the mass-center is
\begin{equation}
\mathbf{R}_{cm}=\sum_{i=1}^4 m_i \mathbf{r}_i/\sum_{i=1}^4 m_i,
\end{equation}
where $m_i$ is the mass of the $i$th quark.

Then the outer product of space and spin wave functions is
\begin{equation}
\Phi_{JJ_z}=[[
\phi^G_{lm}(\mathbf{r})\chi_{s_1m_{s_1}}]_{J_1M_1}
[\psi^G_{LM}(\mathbf{R})\chi_{s_2m_{s_2}}]_{J_2M_2}
]_{J_{12}M_{12}}\varphi^G_{\beta\gamma}(\mathbf{\rho})]_{JJ_z}.
\end{equation}

In GEM, three relative motion wave functions are written as,
\begin{eqnarray}
\phi^G_{lm}(\mathbf{r}) & =& \sum_{n=1}^{n_{max}}c_nN_{nl}r^l
  e^{-\nu_nr^2}Y_{lm}(\hat{\mathbf{r}})\label{gem1} \\
\psi^G_{LM}(\mathbf{R}) & = & \sum_{N=1}^{N_{max}}c_NN_{NL}R^L
  e^{-\zeta_NR^2}Y_{LM}(\hat{\mathbf{R}})\label{gem2} \\
\varphi^G_{\beta\gamma}(\mathbf{\rho}) & = &
\sum_{\alpha=1}^{\alpha_{max}}c_{\alpha}N_{\alpha\beta}\rho^{\beta}
 e^{-\omega_\alpha \rho^2}Y_{\beta\gamma}(\hat{\mathbf{\rho}})\label{gem3}
\end{eqnarray}
Gaussian size parameters are taken as geometric progression
\begin{eqnarray}
\nu_{n}=\frac{1}{r^2_n},& r_n=r_1a^{n-1},&
a=\left(\frac{r_{n_{max}}}{r_1}\right)^{\frac{1}{n_{max}-1}}
\label{geo progress}
\end{eqnarray}
The expression of $\zeta_N,R_N,\omega_\alpha,\rho_\alpha$ in Eqs.
(\ref{gem2}) - (\ref{gem3}) are similar to Eq. (\ref{geo progress}).

The physical state must be in color singlet, which can be constructed
in two ways: color-singlet and color octet,
 \begin{eqnarray}
\left|\xi_1\right\rangle=\left|\mathbf{1}_{12}\otimes\mathbf{1}_{34}\right\rangle,~~~
\left|\xi_2\right\rangle=\left|\mathbf{8}_{12}\otimes\mathbf{8}_{34}\right\rangle.
\label{color}
 \end{eqnarray}
The state in color octet channel is called \textit{\textbf{hidden color}} states by
analogy to states which appear in the nucleon-nucleon problem
\cite{wang0903.4349}.

The total spin of $P\bar P$ and $P\bar V$ and $V \bar P$ system is
only 0, 1 respectively. However, the $V\bar V$ system can coupling
to total spin 0, 1 and 2.

\section{Numerical results and discussion\label{result}}
Solving the Schr\"{o}dinger equation
\begin{eqnarray}
\left(H-E\right)\Psi^{I,I_z}_{J,J_z}=0 \label{schrodinger}
\end{eqnarray}
with Rayleigh-Ritz variational principle, the energies of normal
mesons, $\mathcal{B}\bar{\mathcal{B}}$, $\mathcal{B}\bar{\mathcal{
B}}^*$ and $\mathcal{B^*}\bar{\mathcal{B}}^*$ systems can be
obtained by using different total wave functions, respectively.

To determine if the $\mathcal{B}\bar{\mathcal{B}}$,
$\mathcal{B}\bar{\mathcal{B}}^*$ and $\mathcal{B^*}\bar{\mathcal{
B}}^*$ systems are bound or not, the threshold of the system should
be fixed. Clearly the threshold is governed by two corresponding
meson masses. So one believes that a good fit of meson spectra, with
the same parameters used in four-quark calculations, must be the
most important criterium\cite{weinstein,Bhduri1,Manohar2,Bhduri2,
Brink, Janc1}. Of course, there is another possible threshold for
four-quark system, bottomonium+light meson, e.g., $\Upsilon(1S)+\rho$
for $IJ^P=11^+$ channel. Generally the threshold in this case is lower,
so the bound state $\mathcal{B}^{(*)}\bar{\mathcal{B}}^{(*)}$ will become
a resonance in this channel. Since the transition from
$\mathcal{B}^{(*)}\bar{\mathcal{B}}^{(*)}$ to bottomonium+light meson
involves string rearrangement, we leave this for the future work.

In GEM, the calculated results of normal meson spectra (or the
spectra of $\mathcal{B}\bar{\mathcal{B}}$,
$\mathcal{B}\bar{\mathcal{ B}}^*$ and
$\mathcal{B^*}\bar{\mathcal{B}}^*$ systems) are converged with the
number of gaussians $n_{max}=7~
(n_{max}=7,~N_{max}=7,~\alpha_{max}=12)$, and the size parameter
$r_n~ (r_n,~R_N,~\rho_{\alpha})$ running from $0.1$ to
$2~(2,~2,~6)$~fm. The convergence properties of the energies have
been discussed in detail in Ref.\cite{80114023}. The parameters and
the normal meson spectra in two types of ChQM are listed in Table
\ref{parameters} and \ref{meson spectrum}, respectively.


\begin{table}[htb]
\caption{Parameters of two quark models. The masses of $\pi,\eta$ in
Eqs.(\ref{pi})-(\ref{eta}) are got from the experimental data which
are $m_\pi=0.7$~fm$^{-1}$, $m_{\eta}=2.77$~fm$^{-1}$, respectively;
$m_{\sigma},\Lambda_{\pi}$, $\Lambda_{\eta}$, $\theta_p$ are taken
the same values as Ref.\cite{slamanca}, namely
$m_{\sigma}=3.42$~fm$^{-1}$,
$\Lambda_{\pi}=\Lambda_{\sigma}=4.2$~fm$^{-1}$,
$\Lambda_{\eta}=5.2$~fm$^{-1}$, $\theta_p=-15^o$,
$g^2_{ch}/4\pi$=0.54 }\label{parameters}
\begin{center}
\begin{indented}
\lineup
\item[]\begin{tabular}{@{}llll}\br
& &ChQM1 & ChQM2 \cite{slamanca} \\ \mr
 Quark masses & $m_{u,d}$ (MeV) &313 &313 \\
 &$m_s$(MeV) &525 &555 \\
 &$m_c$(MeV) &1731 &1752\\
 &$m_b$(MeV) &5100 &5100\\\mr
 Confinement &$a_c$(MeV fm$^{-1}$) & 160 & 430\\
 &$\Delta$ (MeV) & -131.1 & 181.1\\
 &$\mu_c$(fm$^{-1}$)&--- &0.7 \\ \mr
 OGE &$\alpha_{0}$ & 2.65 & 2.118\\
 &$r_{0}$(MeV fm) &28.17 &28.17\\
 &$\mu_0$(MeV) &36.976 &36.976\\
 &$\Lambda_0$(fm) &0.075 &0.113\\\br

\end{tabular}
\end{indented}
\end{center}
\end{table}

\begin{table}[htb]
\begin{center}
\caption{Numerical results of normal meson spectrum(unit: MeV)
in the two quark models. The experimental data marked ``Exp." takes
from the latest Particle Data Group\cite{PDG}, and the ground state
of bottom $\eta_b(1s)$ is observed by BABAR Collaboration from the
radiative transition $\Upsilon(3S)\rightarrow \gamma\eta_b$
\cite{Grenier} }\label{meson spectrum}
\begin{indented}
\lineup
\item[]\begin{tabular}{@{}llllllll}
\br
 Meson &ChQM1 & ChQM2 & Exp. & ~~~~Meson & ChQM1 & ChQM2 & Exp. \\ \mr
 $\pi$          &140.1  &153.2 &139.57$\pm$0.00035 &  ~~~~$D_s$     &1966.6 &1991.8 &1968.49$\pm$0.34 \\
 K              &496.2  &484.9 &493.677$\pm$0.016  &  ~~~~$D^*_s$   &2091.1 &2094.1  &2112.3$\pm$0.5\\
 $\rho(770)$    &775.3  &773.1 &775.49$\pm$0.34    &  ~~~~$B^\pm$   &5284.7 &5277.9  &5279.15$\pm$0.31\\
 $K^*(892)$     &917.9  &907.7 &896.00$\pm$0.25    &  ~~~~$B^0$     &5284.7 &5277.9  &5279.53$\pm$0.33\\
 $\omega(782)$  &703.7  &696.6 &782.65$\pm$0.12    &  ~~~~$B^*$     &5324.3 &5318.8  &5325.1$\pm$0.5\\
 $\phi(1020)$   &1016.8 &1011.9 &1019.422$\pm$0.02  &  ~~~~$B^0_s$   &5360.6 &5355.8  &5366.3$\pm$0.6\\
 $\eta_c(1s)$   &2995.7 &2999.8 &2980.3$\pm$1.2     &  ~~~~$B^*_s$   &5403.6 &5400.5  &5412.8$\pm$1.3\\
 $J/\psi(1s)$   &3097.6 &3096.7 &3096.916$\pm$0.011 &  ~~~~$\eta_b(1s)$   &9384.6 &9467.9  &9388.9$^{+3.1}_{-2.3}$(stat)\\
 $D^0$          &1882.2 &1898.4 &1864.84$\pm$0.17   &  ~~~~$\Upsilon(1s)$ &9462.4 &9504.7  &9460.30$\pm$0.26 \\
 $D^*$          &2000.2 &2017.3 &2006.97$\pm$0.19   & \\
\br
\end{tabular}
\end{indented}
\end{center}
\end{table}

By solving the equation (\ref{schrodinger}), the energy of the
$\mathcal{B}\bar{\mathcal{B}}$, $\mathcal{B}\bar{\mathcal{B}}^*$ and
$\mathcal{B}^*\bar{\mathcal{B}}^*$ systems can be obtained. If the
binding energy, $\Delta E =M_{system}-M_{b\bar{q}}-M_{\bar{b}q}$
($q=u,d,s$), is negative, then the system would be bound. According
to the Table. \ref{meson spectrum}, the thresholds of possible
molecular states $\mathcal{B}\bar{\mathcal{B}},
\mathcal{B}\bar{\mathcal{B}}^*,\mathcal{B}^*\bar{\mathcal{B}}^*$ are
easily listed in Table \ref{threshold}.

\begin{table}[htb]
\caption{Threshold energy of $\mathcal{B}\bar{\mathcal{B}},
\mathcal{B}\bar{\mathcal{ B}}^*,\mathcal{B^*}\bar{\mathcal{B}}^*$ in
two quark models.(unit: MeV). }\label{threshold}
\begin{indented}
\lineup
\item[]\begin{tabular}{@{}cccll}
\br
~~~J~~~ &~~~I~~~ & $M_1M_2$  & E$_{th}$(ChQM1) & E$_{th}$(ChQM2) \\
\mr
&0,1             &$\bar{B}^0B^+$                  &10569.4   &10555.8   \\
0&0              &$B_s^-B_s^+$                    &10721.2   &10711.6  \\
&$\frac{1}{2}$   &$B_s^-B^0/B_s^+B^-$             &10645.3   &10633.7  \\
\mr

&0,1            &$\bar{B}^0B^{*+}$                &10609     &10596.7 \\
1&0             &$B_s^-B_s^{*+}$                  &10764.2   &10756.3 \\
 &$\frac{1}{2}$ &$B_s^-B^{*0}/B_s^+B^{*-}$        &10684.9   &10674.6 \\
 &              &$B_s^{*-}B^{0}/B_s^{*+}B^{-}$    &10688.3   &10678.4 \\ \mr
&0,1            &$\bar{B}^{*0}B^{*+}$             &10648.6   &10637.6  \\
0,1,2&0         & $B_s^{*-}B_s^{*+}$              &10807.2   &10801 \\
&$\frac{1}{2}$  &$B_s^{*-}B^{*0}/B_s^{*+}B^{*-}$  &10727.9   &10719.3 \\
\br
\end{tabular}
\end{indented}
\end{table}

The color-singlet, color-octet channel, and channel coupling
calculation of $\mathcal{B}\bar{\mathcal{B}},
\mathcal{B}\bar{\mathcal{ B}}^*,\mathcal{B^*}\bar{\mathcal{B}}^*$
systems are done in the two types of ChQM, and the results are
presented in Table \ref{PPbb}-\ref{VVbb}, in which the results are
denoted by ``$1\otimes1$", ``$8\otimes8$" and ``channel coupling",
respectively. Due to the strong interaction is invariant under the
rotation of isospin, the different states corresponding to the
different components of isospin $I$ are degenerate, so we present
the results for each total isospin $I$. From Table \ref{PPbb}-\ref{VVbb},
we can see two models give very similar results.

In the S-wave $P\bar{P}$ system, the quantum numbers $J^{P}$ are
$0^+$. Apart from scalar meson $\sigma$, the pseudoscalar mesons
e.g. $\pi,K,\eta$ cannot be exchange in
$\mathcal{B}\bar{\mathcal{B}}$ system because of the parity
conservation. One can find in the Table \ref{PPbb} that the $\sigma$
meson exchange do not contribute enough attraction to bind the
$\mathcal{B}\bar{\mathcal{B}}$ system in color-singlet channel
in the ChQM1. Due to the $\sigma$-exchange also occurs between $ss$
pair and $us$ or $ds$ pairs, loosely bound states of
$\mathcal{B}\bar{\mathcal{B}}$ with $I=0(s)$, or $\frac{1}{2}$ are
obtained in the ChQM2.  Noteworthily, there is no one-gluon-exchange
between two separate mesons just in this channel. More bound states
are formed if we take the coupling of color-singlet and color-octet
channels into account. Obviously, in the color octet channel, the
color-magnetic terms of OGE between two separate colorful mesons
contribute attraction to $\mathcal{B}\bar{\mathcal{B}}$ system,
because of the requirement of total color-singlet of the state. In
this case, due to the masses of $u,~d$ quarks are much smaller than
the mass of $b$ quark, the cross matrix between color-singlet and
-octet channels of the color-magnetic interaction, which is in
proportional to $1/(m_i m_j)$, is very large. So the energy of each
$\mathcal{B}\bar{\mathcal{B}}$ system is depressed by it, which was
discussed in detail in Ref. \cite{weinstein, PRD81114025}.

\begin{table}[ht]
\caption{The binding energy (unit: MeV) of
$\mathcal{B}\bar{\mathcal{B}}$. The "$1\otimes1$", "$8\otimes8$" and
"channel coupling" represent $\mathcal{B}\bar{\mathcal{B}}$ in
color-singlet, color-octet and coupling of color-singlet and
color-octet channel, respectively. }\label{PPbb}
\begin{indented}
\lineup
\item[]\begin{tabular}{@{}lccc|ccc}
\br & & ChQM1& & & ChQM2&\\
Isospin &$1\otimes1$ &$8\otimes8$ & channel coupling &$1\otimes1$ &$8\otimes8$ & channel coupling\\
 \mr
I=$\frac{1}{2}$&0.5 &51.5 &-17.5 &-0.2 &97.7 &-2.6 \\
I=1            &0   &0.1  &-72.6 &0 &45.6 &-29.9  \\
I=0(\textit{l})&0   &0.1  &-72.6 &0 &45.6 &-29.9  \\
I=0(\textit{s})&0.3 &77.4 &0.2   &-0.7 &137.5 &-1.5  \\\br
\end{tabular}
\end{indented}
\end{table}

For the $\mathcal{B}\bar{\mathcal{B}}^*$ system, the calculation
results are listed in Table \ref{VPbb}. The $\sigma,\pi,\eta$ mesons
can all be exchanged in such systems. Two bound states are both
found in two quark models in color-singlet single channel. For $I=1$
state, the binding energy are both about $-1$ MeV with regard to
$\mathcal{B}\bar{\mathcal{B}}^*$ threshold in two quark models, and
the distance of each pair quarks in ChQM1 are
\begin{eqnarray}
\sqrt{\langle r^2_{12}\rangle}=0.63~\mbox{fm},& & \sqrt{\langle
r^2_{34}\rangle}=0.63~\mbox{fm},
\nonumber \\
\sqrt{\langle r^2_{13}\rangle}=2.1~\mbox{fm},& & \sqrt{\langle
r^2_{24}\rangle}=2.1~\mbox{fm},
\nonumber \\
\sqrt{\langle r^2_{14}\rangle}=2.02~\mbox{fm}, & & \sqrt{\langle
r^2_{23}\rangle}=2.18~\mbox{fm}. \nonumber
\end{eqnarray}
The distance between quark and quark or antiquark belonging to
different mesons are larger than that between quark and antiquark
belonging to the same meson, the results show a clear molecule
structure. It is reasonable to interpret the state $Z_b(10610)$
reported by Belle collaboration as the molecular state $B\bar{B}^*$
with $I(J^{PC})=1(1^{+-})$. For this state, there are also other
thresholds, $\Upsilon(1S)\rho$ and $h_b\pi$ \cite{Cleven:2011gp}.
The energy of the state is a little higher than these thresholds.
Because of the different color structures the states $B\bar{B}^*$
and $\Upsilon(1S)\rho$, $h_b\pi$ have, the transition involves the
color structure rearrangement, $B\bar{B}^*$ may appear as a
resonance in the $\Upsilon(1S)\rho$ and $h_b\pi$
channels~\cite{benezene}. The calculation of the transition, which
is out of the scope of the present work, is left for future work. In
our calculation, the $I=0$ state without strange quark is also a
bound state with binding energy $-12.1$ MeV in two quark models. In
the ChQM2, in addition to the above two states, the
$I=\frac{1}{2}$ and $0(s)$, also form a bound state for the
$\sigma$-exchange contributes to these channels. The color-singlet
singlet and hidden color channel coupling leads to that all the
states are bound.
\begin{table}[htb]
\caption{The binding energy (unit: MeV) of
$\mathcal{B}\bar{\mathcal{B}}^*$. }\label{VPbb}
\begin{indented}
\lineup
\item[]\begin{tabular}{@{}clccc|ccc}
\br
& & & ChQM1& & & ChQM2&\\
$J^{P}$&Isospin &$1\otimes1$ &$8\otimes8$ & channel coupling &$1\otimes1$ &$8\otimes8$ & channel coupling\\
\mr
&I=$\frac{1}{2}$  &0.4  &-21.2   &-89.8    &-0.2 &28.9 &-40.9  \\
$1^{+}$&I=1      &-1.3   &-88.1   &-164.3  &-1.1 &-35.5 &-107.4  \\
&I=0(\textit{l})  &-12.1 &-47.2   &-122.6  &-12.1 &2.0 &-69.8  \\
&I=0(\textit{s})  &0.35  &14.8   &-47.3    &-1.2 &77.5 &-3.9  \\ \br
\end{tabular}
\end{indented}
\end{table}

\begin{table}[ht]
\caption{The binding energy (unit: MeV) of
$\mathcal{B}^*\bar{\mathcal{B}}^*$. }\label{VVbb}
\begin{indented}
\lineup
\item[]\begin{tabular}{@{}clccc|ccc}
\br
& & & ChQM1& & & ChQM2&\\
$J^{P}$&Isospin &$1\otimes1$ &$8\otimes8$ & channel coupling &$1\otimes1$ &$8\otimes8$ & channel coupling\\
\mr
&I=$\frac{1}{2}$ &0.4  &-111.8 &-172.9  &-0.5 &-56.6 &-108.4 \\
$0^{+}$  &I=1   &-3.4 &-194.4 &-266.2   &-3.0 &-132.3 &-195.5 \\
&I=0(\textit{l}) &0.4  &-104.9 &-175.9  &0.4 &-49.8 &-115.3 \\
&I=0(\textit{s}) &0.3  &-58.9  &-112.2  &0 &7.6 &-37.7 \\ \mr
&I=$\frac{1}{2}$ &0.4  &-71.4  &-129.0  &-0.3 &-22.5 &-73.9 \\
$1^{+}$&I=1     &-1.2 &-135.7 &-201.5   &-0.9 &-83.2 &-144.6 \\
&I=0(\textit{l}) &0.4  &-94    &-160.1  &0.5 &-44.9 &-107.6 \\
&I=0(\textit{s}) &0.3  &-34.9  &-86.0   &-0.2 &26.5 &-21.2 \\ \mr
&I=$\frac{1}{2}$ &0.4  &-2.6   &-56.4   &0 &36.1 &-19.3 \\
$2^{+}$&I=1     &0.3  &-37.6  &-97.0    &0.4 &-1.9 &-62.6 \\
&I=0(\textit{l}) &-11.1&-74.3  &-133.0  &-11.0 &-35.7 &-95.1 \\
&I=0(\textit{s}) &0.3  &8.4    &-41.0   &-1.1 &61.1 &-5.7 \\
 \br
\end{tabular}
\end{indented}
\end{table}

The S-wave $\mathcal{B^*}\bar{\mathcal{B}^*}$ systems have quantum
numbers $J^{PC}=0^{++},~1^{+-}$, and $2^{++}$ for the neutral
states. The $\pi,~\eta,~\sigma$ mesons can all be exchanged for
I=0(\textit{l}) and I=1, while only $\eta$ is exchangeable for
I=0(\textit{s}) and $\frac{1}{2}$ states. The $\sigma$ interaction
is always attractive between the lightest $u,~d$ quarks. According
to Eq. (\ref{pi}), the $\pi$-exchange is attractive for the states
with $I(J^{P})=1(0^{+}),~1(1^{+}),~0(2^{+})$ and makes these states
are all bound states, which are shown in Table \ref{VVbb} in the
color-singlet single channel calculation. The
$\mathcal{B}^*\bar{\mathcal{B}}^*$ with $I(J^{P})=1(1^{+})$ has
binding energy about -1 MeV in two quark models, and the distance of
each pair quarks are
\begin{eqnarray}
\sqrt{\langle r^2_{12}\rangle}= 0.64~\mbox{fm},& & \sqrt{\langle
r^2_{34}\rangle}= 0.64~\mbox{fm},
\nonumber \\
\sqrt{\langle r^2_{13}\rangle}=2.2~\mbox{fm},& & \sqrt{\langle
r^2_{24}\rangle}=2.2~\mbox{fm},
\nonumber \\
\sqrt{\langle r^2_{14}\rangle}=2.11~\mbox{fm}, & & \sqrt{\langle
r^2_{23}\rangle}=2.27~\mbox{fm}. \nonumber
\end{eqnarray}
The assignment of the newly observed state $Z_b(10650)$ to a
molecular state $\mathcal{B}^*\bar{\mathcal{B}}^*$ with
$I(J^{PC})=1(1^{+-})$ is favored. In the hidden color channel,
almost all the states, except the one with $IJ^P=02^+$ and hidden
strange, are become bound. Again the channel coupling between
color-singlet and hidden color channels makes all the states bound.
In the present work, the $S$-$D$ mixing of $Z_b$ is not taken into
account. the mixing will be important for states with energy on the
threshold. From the calculation of deuteron, we estimate the
$S$-$D$ mixing will increase the binding energy of $Z_b$ about 2 MeV.

\section{Summary}
In the framework of chiral quark model, a systematical study of the
mass spectra of $\mathcal{B}\bar{\mathcal{B}},~
\mathcal{B}\bar{\mathcal{B}}^*$ and
$\mathcal{B}^*\bar{\mathcal{B}}^*$ systems is performed. The states
$\mathcal{B}\bar{\mathcal{B}}^*$ and
$\mathcal{B}^*\bar{\mathcal{B}}^*$ with quantum numbers $I(J^{PC})=1(1^{+-})$ are
shown to be bound, which are respectively good candidates for the
charged bottomonium-like resonances $Z_b(10610)$ and $Z_b(10650)$
newly observed by Belle collaboration. The color-singlet single
channel calculation also shows that the states $B\bar{B}^*$ with
$I(J^{PC})=0(1^{++})$, $B^*\bar{B}^*$ with
$I(J^{PC})=1(0^{++}),~0(2^{++})$ are bound states with a few MeV
binding energy.

Recently Belle collaboration reported their high precision
measurement of bottomonium mass: $M[\Upsilon(5S)]=$10.87 GeV
\cite{prl100112001}. If the molecular states $B^*\bar{B}^*$ with
$I(J^{P})=1(0^{+})$ really exist, it could be observed in final
state $\Upsilon(1S)\rho$ at the Belle, BaBar, LHC and other
collaborations. Due to the phase space limitation, the isoscalar
states $B\bar{B}^*(J^{P}=1^{+})$ and $B^*\bar{B}^*(J^{P}=2^{+})$ may
be observed in decays of excited bottomonium which above the
$\Upsilon(5S)$.

The $\sigma$-exchange plays important role for binding the
$\mathcal{B}\bar{\mathcal{B}},~ \mathcal{B}\bar{\mathcal{B}}^*$ and
$\mathcal{B}^*\bar{\mathcal{B}}^*$  with $I=\frac{1}{2}$ and $0(s)$
in the ChQM2. To search these molecular states in the future
experiment will test the contribution of the $\sigma$-exchange in
the chiral constituent quark model.

The hidden color channel effect is complicated in multi-quark
systems. Here we extend directly the quark model for
colorless cluster to the colorful cluster in the study of
$\mathcal{B}\bar{\mathcal{B}},~\mathcal{B}\bar{\mathcal{B}}^*$ and
$\mathcal{B}^*\bar{\mathcal{B}}^*$ systems. In our calculation, the
color-octet channel plays a dominate role in producing deeply bound
states. If the quark-antiquark interaction in color singlet can be
extended directly to color-octet by Casimir scaling~\cite{Bali},
then the OGE interaction will be attractive between two color-octet
cluster in some multi-quark systems. So it is inevitably to produce
deeply bound states for hidden-bottom states because of the too
small kinetic energy. More experimental data on bottomonium-like
resonances are needed to check the Casimir scaling, and cast doubt
on the validity of applying the chiral constituent quark model to
the hidden color channel directly.

\ack{This work is supported partly by the National
Science Foundation of China under Contract Nos. 11047023, 11035006,
11175088, 11047140, and the Science Foundation of Guizhou Provincial
Eduction Department under Grant No. 20100084, and the Science
Foundation of Guizhou Science and Technology Department under Grant
No. J[2011]2364, and the key support discipline of Guizhou province
No. [2011]275.}

\end{document}